\documentclass[pra,preprint,amsmath,amssymb,showpacs]{revtex4}
\usepackage{bm}
\usepackage{graphics}

\newcommand{\vx}{\vec{x}\,}
\begin{document}

\title{Rayleigh functional for nonlinear systems }
\author{Valery S. Shchesnovich}
\email{valery@loqnl.ufal.br}
\author{ Solange B. Cavalcanti}
 \affiliation{ Departamento de F\'{\i}sica - Universidade
Federal de Alagoas, Macei\'o AL 57072-970, Brazil }

\begin{abstract}
We introduce Rayleigh functional for nonlinear systems. It is
defined using the energy functional and the normalization
properties of the variables of variation. The key property of the
Rayleigh quotient for linear systems is preserved in our
definition: the extremals of the Rayleigh functional coincide with
the stationary solutions of the Euler-Lagrange equation. Moreover,
the second variation of the Rayleigh functional defines stability
of the solution.  This gives rise to a powerful numerical
optimization method in the search for the energy minimizers.   It
is shown that the well-known imaginary time relaxation is a
special case of our method. To illustrate the method we find the
stationary states of Bose-Einstein condensates in various
geometries. Finally, we show that the Rayleigh functional also
provides a simple way to derive analytical identities satisfied by
the stationary solutions of the critical nonlinear equations.

\bigskip
\end{abstract}
\maketitle

\section{Introduction}
\label{intro}

In dealing with  systems of equations having  a large number of
degrees of freedom one frequently adopts an approximation which
leads to a nonlinear partial differential equation.  For instance,
the mean-field approach in the statistical physics  is based on
the introduction of an order parameter governed by a nonlinear
equation. For example, the mean-field theory of the Bose-Einstein
condensate of a degenerate quantum gas is based on the
Gross-Pitaevskii equation \cite{BEC}. The Gross-Pitaevskii
equation is the nonlinear Schr\"odinger equation  with an external
potential. It describes the so-called ``matter waves'' -- the
nonlinear collective modes of the degenerate quantum gas below the
condensation temperature. In many other cases the approximate
nonlinear equations, appearing as the leading order description of
nonlinear waves in various branches of modern physics, for
instance, in nonlinear optics, plasma, and on water, are of the
nonlinear Schr\"odinger class (see, for instance, Ref. \cite{NLS}
and the references therein). In physics one is especially
interested in the stationary solutions (a.k.a. the stationary
points) of the governing equations, in particular, in the
stationary points which minimize the energy. Only in some
exceptional cases the solution can be obtained analytically and
one has to rely on numerical simulations. If the nonlinear
equation in question possesses the Lagrangian formulation  then
the stationary points can be found by a nonlinear optimization
method.

We propose  a new optimization method for nonlinear partial
differential equations, in particular, for the equations of the
nonlinear Schr\"odinger (NLS) type. The method consists of a
numerical minimization of the Rayleigh functional. In a broader
perspective, we introduce the concept of Rayleigh functional for
nonlinear systems and show how it can be employed in both
numerical and analytical analysis of equations of physical
importance.

In our case, the optimization problem is to find the ground state
of a nonlinear system, i.e. the stationary solution which
minimizes the energy. We work in the space of smooth functions
$\psi(\vec{x},t)$ of $n$ spatial variables $\vec{x} =
(x_1,...,x_n)$ and time variable, which have bounded $l_2$-norm:
$||\psi||^2 \equiv \int \mathrm{d}^n\vec{x} |\psi|^2 < \infty$.
Thus we consider only bounded stationary solutions which decay to
zero as $x_k\to\infty$ and correspond to finite energies.

We adopt the following energy functional:
\begin{equation}
E\{\psi,\psi^*\} = \int\mathrm{d}^n\vx\mathcal{E}
(\vx,\psi(\vx,t),\psi^*(\vx,t),\nabla\psi(\vx,t),\nabla\psi^*(\vx,t))
\label{EQ1}\end{equation}
and  assume that it has the phase invariance symmetry $\psi \to
e^{i\theta}\psi$ (we use the complex conjugate functions $\psi$
and $\psi^*$ as independent variational variables instead of the
real and imaginary parts of $\psi$). The phase invariance symmetry
results in conservation of the $l_2$-norm.

Our method also works for the generalization of the energy
functional (\ref{EQ1}) to several variables of variation:
$\psi_k(\vx)$, $k=1,\ldots,m$ (this case is discussed below), and
to the higher order derivatives: $\nabla^p\psi$, $p=1,\ldots,s$.
We adopt the notations used in statistical physics, an important
area for applications. Hence, the dependent variable $\psi(\vx,t)$
will be referred to as the order parameter.

The well known Gross-Pitaevskii functional for the order parameter
of a Bose-Einstein condensate,
\begin{equation}
E = \int\mathrm{d}^3\vx
\left\{\frac{\hbar^2}{2m}|\nabla\psi(\vx,t)|^2 +
V(\vx)|\psi(\vx,t)|^2 + \frac{gN}{2}|\psi(\vx,t)|^4\right\},
 \label{EQ2}\end{equation}
belongs to the class specified by equation (\ref{EQ1}). Here
$\vec{x} = (x,y,z)$, $N$ is the number of atoms in the condensate,
$g = 4\pi\hbar^2a_s/m$ is the interaction coefficient due to the
s-wave scattering of the atoms and $V(\vx)$ is the  trap potential
(created by a magnetic field or non-resonant laser beams). The
order parameter is normalized to 1 in equation (\ref{EQ2}).

Let us briefly recall the basics of the nonlinear optimization.
One is interested in the  stationary state, given as $\psi(\vx,t)
= e^{-i\mu t}\Psi(\vx)$, where $\mu$ is the chemical potential.
The Euler-Lagrange equation corresponding to the energy functional
(\ref{EQ1})  reads
\begin{equation}
i\frac{\partial\psi}{\partial t} = \frac{\delta E}{\delta \psi^*}
\equiv \frac{\partial \mathcal{E}}{\partial \psi^*} -
\nabla\frac{\partial \mathcal{E}}{\partial \nabla\psi^*}.
 \label{EQ3}\end{equation}
Here (and below) we denote $\delta F/\delta\psi^*$  the
variational derivative of a functional $F$ with respect to
$\psi^*$ (thus we part with the usual notation of the latter by
using the symbol $\nabla$, while $\nabla$ is reserved for the
usual gradient of a function). The stationary state satisfies the
equation $\mu\Psi = {\delta E}/{\delta \Psi^*}$. The idea of the
imaginary time relaxation method is based on the fact that the
variational derivative of a functional is an analog of the
gradient of a function. Thus, by introducing the ``imaginary
time'' $\tau = it$ in equation (\ref{EQ3}) one forces the order
parameter to evolve in the direction of maximum decrease in the
energy. The attractor of such evolution is, hopefully, a local
minimum (in general, just a stationary point). The imaginary time
evolution does not conserve the $l_2$-norm and  one must normalize
the order parameter directly during  the evolution. In other
words, one allows the order parameter to evolve in the space of
arbitrary functions but normalizes the solution after each step by
the prescription $\psi\to\frac{\psi}{||\psi||}$. This can be
formulated in a single equation:
\begin{equation}
\left.\frac{\partial \psi}{\partial \tau} = - \frac{\delta
E\{f,f^*\}}{\delta f^*}\right|_{f = \frac{\psi}{||\psi||}}.
 \label{EQ4}\end{equation}

The method of Lagrange multipliers can be used to take into
account the constraints imposed  on the variables of variation
(for instance, the fixed norm). It consists of a numerical
minimization of an augmented functional which is the energy
functional plus the constraint  with a Lagrange multiplier. The
combined functional has the stationary solution as its extremal
point and the constrained minimization problem is  converted into
an unconstrained one.  In this case, the time evolution of
equation (\ref{EQ4}) can be substituted by a finite-step
minimization scheme, such as the method of steepest descent or the
conjugate gradient method, supplemented by  an appropriate
line-search algorithm (consult, for instance, Refs.
\cite{OPTIM1,OPTIM2}). For instance, the following two functionals
are used
\begin{equation}
F_1\{\psi,\psi^*\} = E\{\psi,\psi^*\} - \mu
\int\mathrm{d}^n\vx|\psi|^2,\quad F_2\{\psi,\psi^*\} =
E\{\psi,\psi^*\} + \frac{1}{2}\left(\lambda -
\int\mathrm{d}^n\vx|\psi|^2\right)^2,
 \label{EQ5}\end{equation}
where the variable of variation is arbitrary (has arbitrary
$l_2$-norm). Indeed, the first of these functionals evidently has
the stationary solutions as its extremals, while the second has
the variation ${\delta E}/{\delta \psi^*} -
(||\psi||^2-\lambda)\psi$. Setting $\mu = (\lambda - ||\psi||^2)$
we get the stationary point by equating the variation of $F_2$ to
zero. The use of the functional $F_2$ in the search for the energy
minimizers was advocated recently in Ref.~\cite{SobolGrad}. The
point is that the trivial solution $\psi = 0$, being an extremal,
frequently makes reaching other solutions difficult with the use
of the functional $F_1$, while functional $F_2$ is free from such
a flaw.

 The simplest numerical realization of the
minimization method is given by the steepest descent scheme:
\begin{equation}
\psi_{k+1} = \psi_k - \beta_{k}\frac{\delta
F\{\psi_k,\psi_k^*\}}{\delta \psi_k^*},
\label{EQ6}\end{equation}
where the parameter $\beta_k$ is selected by an appropriate
line-search algorithm.

This paper is organized as follows. In section \ref{Ray} we
introduce the  Rayleigh functional for the energy given by
equation (\ref{EQ1})  and discuss its properties as a variational
functional. The generalization to several order parameters is also
considered. In section \ref{Num} we present  examples of the
nonlinear optimization based on the Rayleigh functional. As an
application of the Rayleigh functional in the analytical approach,
in section \ref{criticalNLS} we relate an identity satisfied by
the stationary solutions of the  critical NLS equations to the
scale invariance symmetry of the Rayleigh functional. Finally, in
section \ref{Concl} the advantage of the numerical optimization
method based on the Rayleigh functional is discussed.

\section{Definition and properties of the Rayleigh functional}
\label{Ray}

Before formulating our nonlinear optimization method, it would be
instructive to recall how the eigenfunctions of a linear operator
can be obtained numerically. Consider, for instance, the textbook
problem of finding the eigenvalues and eigenfunctions of the
Hamiltonian operator $\hat{H} = -\nabla^2 + V(\vx)$  for a quantum
particle in a potential well (we use $\hbar = 1$ and $m = 1/2$).
This problem can be reformulated as an optimization problem by
employing the well-known Rayleigh quotient \cite{LanLif}
\begin{equation}
\mathcal{R}\{\psi,\psi^*\} =
\frac{\int\mathrm{d}^3\vx\psi^*(\vx)\hat{H}\psi(\vx)}
{\int\mathrm{d}^3\vx|\psi(\vx)|^2} =
\int\mathrm{d}^3\vx\frac{\psi^*(\vx)}{||\psi||}\hat{H}\frac{\psi(\vx)}{||\psi||}.
\label{EQ7}\end{equation}
Here the function of variation $\psi$ is arbitrary, i.e. not
normalized. The eigenfunctions are the stationary points of the
Rayleigh quotient. Indeed
\begin{equation}
\frac{\delta \mathcal{R}}{\delta \psi^*} =
\frac{1}{||\psi||^2}\left(\hat{H}\psi - \varepsilon\psi\right)= 0,
\label{EQ8}\end{equation}
where the eigenvalue is given as $\varepsilon =
\mathcal{R}\{\psi,\psi^*\}$, i.e. it is the value of $\mathcal{R}$
at the eigenfunction.

Note that the Rayleigh quotient allows one to cast the constrained
minimization problem (with the constraint $||\psi|| = 1$) into an
unconstrained one.

\subsection{Rayleigh functional for a single order parameter}
\label{scalar}

Stationary solutions of nonlinear systems, the so-called nonlinear
modes,  can be considered as the generalization of the
eigenfunctions. They can be found in a way similar to the solution
of the above eigenvalue problem.

In contrast to the linear systems,  a \textit{nonlinear} one
usually possesses   continuous families of stationary solutions,
each solution corresponding to a different chemical potential (an
analog of the energy level in the nonlinear case). In general, the
chemical potential  takes its values from an infinite interval of
the real line.  By fixing the $l_2$-norm we introduce it as a
parameter in the energy functional. Thus, the chemical potential,
being a continuous function of the norm, is also fixed (in
general, we get away with a finite number of \textit{distinct}
values lying on the different branches of the function $\mu =
\mu(||\psi||)$). The only exception is the so-called critical case
when a continuous family of  stationary solutions corresponds to
the same $l_2$-norm (see also section \ref{criticalNLS}, where we
discuss the critical case). Therefore, apart from the critical
case, by fixing the norm we select a finite number of solutions.
Among them there is the energy minimizer (for a given value of the
conserved $l_2$-norm). We can reconstruct the continuous family of
the stationary solutions by using different values of the norm. In
the case of Bose-Einstein condensates, for instance, this
corresponds to fixing the number of atoms and looking for the
corresponding ground state solution.

It is assumed that the nonlinear equation in question has a phase
invariance resulting in the $l_2$-norm conservation. In the
following we will distinguish between the normalized and
not-normalized functions, denoting the former by $f(\vec{x},t)$
and the latter by $\psi(\vec{x},t)$. For convenience, we set the
$l_2$-norm of the stationary solution equal to 1, $||f|| = 1$.
This normalization is performed by using a scale transformation of
the function of variation and results in the explicit appearance
of the value of $l_2$-norm in the energy functional (note that the
chemical potential may also be scaled appropriately, as in
equations (\ref{EQincoh}) and (\ref{EQcoh}) of section \ref{Num}).
For example, in equation (\ref{EQ2}) we have the coefficient $N$
at the nonlinear term, after we set the $l_2$-norm of the solution
to 1, and the value of the energy functional is, in fact, the
energy of the condensate per atom.

For the energy given by equation (\ref{EQ1}) the following
functional can serve as the nonlinear generalization of the
Rayleigh quotient:
\begin{equation}
\mathcal{R}\{\psi,\psi^*\} =
E\{f,f^*\}|_{f=\frac{\psi}{||\psi|||}} =
\int\mathrm{d}^n\vx\mathcal{E}
\left(\vx,\frac{\psi(\vx,t)}{||\psi||},\frac{\psi^*(\vx,t)}
{||\psi||},\frac{\nabla\psi(\vx,t)}{||\psi||},
\frac{\nabla\psi^*(\vx,t)}{||\psi||}\right).
\label{EQ9}\end{equation}
Here the $l_2$-norm $||\psi||$ is not fixed but is a functional of
$\psi$ and $\psi^*$ (since the norm is constant \textit{function}
of $\vx$ we can take it out of the gradient operator $\nabla$).
Note that $\mathcal{R}$ is a compound functional: it depends on
the complex conjugate functions $\psi(\vx)$ and $\psi^*(\vx)$
through the normalized ones, $f(\vx)$ and $f^*(\vx)$, used in the
energy functional $E$. This simple functional turns out to be very
helpful in the search for the energy minimizers.

Let us discuss the properties of  functional (\ref{EQ9}). First,
its extremals  are  stationary points of the corresponding
nonlinear equation, equation (\ref{EQ3}), similar as in the case
of the Rayleigh quotient (\ref{EQ7}) and equation (\ref{EQ8}).
(The unique stationary point is selected by the value of the
$l_2$-norm, appearing explicitly in the energy functional as a
\textit{parameter}). This follows from a simple calculation:
\[
\delta\mathcal{R} = \int\mathrm{d}^n\vx\left\{\frac{\delta
E\{f,f^*\}}{\delta f^*}\delta f^* + \frac{\delta
E\{f,f^*\}}{\delta f}\delta
f\right\}\qquad\qquad\qquad\qquad\qquad
\]
\[
= \int\mathrm{d}^n\vx\left\{ \frac{\delta E\{f,f^*\}}{\delta
f^*}\frac{1}{||\psi||}\left(\delta\psi^* -
\frac{\psi^*}{2||\psi||^2}
\int\mathrm{d}^n\vx^\prime(\psi\delta\psi^*+\psi^*\delta\psi)\right)
\right.
\]
\begin{equation}
\qquad \qquad\left. + \frac{\delta E\{f,f^*\}}{\delta
f}\frac{1}{||\psi||}\left(\delta\psi -\frac{\psi}{2||\psi||^2}
\int\mathrm{d}^n\vx^\prime
(\psi\delta\psi^*+\psi^*\delta\psi)\right)
\right\}
\label{EQ10}\end{equation}
where we have used that
\[
\delta f = \frac{1}{||\psi||}\left\{\delta \psi -
\frac{\psi}{2||\psi||^2}\int\mathrm{d}^n
\vx(\psi\delta\psi^*+\psi^*\delta\psi)\right\}.
\]
By interchanging the order of integration in the double integral
in formula (\ref{EQ10}) and gathering  the terms with
$\delta\psi^*$ we obtain the variational derivative
\begin{equation}
\frac{\delta \mathcal{R}\{\psi,\psi^*\}}{\delta \psi^*} = \left.
\frac{1}{||\psi||}\left\{\frac{\delta E\{f,f^*\}}{\delta
f^*}-\mathrm{Re}\left(\int\mathrm{d}^n\vx\frac{\delta
E\{f,f^*\}}{\delta f^*}f^*\right)f\right\}\right|_{f =
\frac{\psi}{||\psi||}}.
\label{EQ11}\end{equation}
From equation (\ref{EQ11}) it is quite clear that the stationary
points of equation (\ref{EQ3}) are  extremals of the Rayleigh
functional and vice versa. The corresponding chemical potential is
equal to the integral in the second term on the r.h.s. of equation
(\ref{EQ11}), i.e.
\begin{equation}
\mu = \mathrm{Re}\left(\int\mathrm{d}^n\vx\frac{\delta
E\{f,f^*\}}{\delta f^*}f^*\right),
\label{EQ12}\end{equation}
which fact can be  verified by direct integration of equation
(\ref{EQ3}), i.e. $\mu f =  \frac{\delta E\{f,f^*\}}{\delta f^*}$,
and taking into account the normalization of $f$. Equation
(\ref{EQ12}) plays here the role of the expression for the energy
level $\varepsilon$ in the above problem of a quantum particle.

The Rayleigh functional also distinguishes the local minima among
the stationary points. To see this let us compute its second
variation. Assuming that $\psi$ is  a stationary point, we get
\[
\delta^2\mathcal{R}\{\psi,\psi^*\} = \left(\delta^2E\{f,f^*\}
\bigr|_{\delta^2f=0} +\int\mathrm{d}^n\vx\left\{\frac{\delta
E\{f,f^*\}}{\delta f^*}\delta^2f^* + \frac{\delta
E\{f,f^*\}}{\delta f}\delta^2f\right\}\right)\biggr|_{f =
\frac{\psi}{||\psi||}}
\]
\[
=\left(\delta^2E\{f,f^*\}\bigr|_{\delta^2 f = 0} +
\int\mathrm{d}^n\vx\left\{\mu f\delta^2f^* + \mu
f^*\delta^2f\right\}\right)\biggr|_{f = \frac{\psi}{||\psi||}},
\]
where we have used the fact that $f$ satisfies the stationary
equation $\mu f = \frac{\delta E\{f,f^*\}}{\delta f^*}$.
Integrating by parts and taking into account the normalization of
$f$, i.e. $\delta||f||^2 = \delta \int\mathrm{d}^n\vx|f|^2 = 0$,
we arrive at
\begin{equation}
\delta^2\mathcal{R}\{\psi,\psi^*\} =
\left(\delta^2E\{f,f^*\}\bigr|_{\delta^2 f = 0} -
2\mu\int\mathrm{d}^n\vx|\delta f|^2\right)\biggr|_{f =
\frac{\psi}{||\psi||}}. \label{EQ13}\end{equation} The subscript
in the first term on the r.h.s. of this equation means that the
variable of variation of the energy functional is, in fact,
$f(\vx)$. Taking this into account, one immediately recognizes on
the r.h.s. of equation (\ref{EQ13}) the second variation of the
functional $F_1\{f,f^*\}$ defined in equation (\ref{EQ5}) and
evaluated in the space of normalized functions. Therefore, the
minima of the Lagrange modified energy functional (for a fixed
nonzero $l_2$-norm) are also minima of the Rayleigh functional and
vice versa. Importantly,  the Rayleigh functional does not contain
the trivial solution $\psi=0$ among its extremals, in contrast to
the Lagrange modified energy functional.

Finally, let us discuss the functional of Ref. \cite{GRQ}, which
was proposed as  another possible generalization of the Rayleigh
quotient to the equations of the nonlinear Schr\"odinger type. In
the latter work in the numerical search for the stationary
solutions of two-dimensional Gross-Pitaevskii equation the
following functional was employed
\begin{equation}
F\{\psi,\psi^*\} = \frac{\int\mathrm{d}\vx^2 \psi^*(-\nabla^2
-\lambda +\vx^2 + U|\psi|^2)\psi}{\int\mathrm{d}\vx^2|\psi|^2}.
\label{cras}\end{equation} This functional  can be expressed as
follows
\begin{equation}
F\{\psi,\psi^*\} = \mathcal{R}\{\psi,\psi^*\} - \lambda +
\frac{U\int|\psi|^4}{||\psi||^2} - \frac{NU\int
|\psi|^4}{||\psi||^4}
\label{EQnull}\end{equation}
Here the Rayleigh functional is given by $\mathcal{R} =
\int\mathrm{d}\vx^2f^*(-\nabla^2 +\vx^2 + NU|f|^2)f$ with $f =
\psi/||\psi||$ and  $N$ is the number of atoms corresponding to
the  stationary state. Note that the first variation of the
Rayleigh functional is zero on a stationary solution and the last
two terms in formula (\ref{EQnull})  have nonzero variation unless
the norm $||\psi||$ is kept fixed ($||\psi||^2 = N$). Hence, the
functional $F\{\psi,\psi^*\}$ introduced in Ref. \cite{GRQ} cannot
be employed for \textit{unconstrained} minimization.

Now let us show that, in fact, the Euler-Lagrange equation for the
functional $F\{\psi,\psi^*\}$ defined by equation (\ref{cras}) is
self-contradictory (in other words, the first variation of
$F\{\psi,\psi^*\}$ is never zero) and, hence, does not lead to any
stationary solutions at all. The Euler-Lagrange equation for the
functional (\ref{cras}) reads
\[
(-\nabla^2 + \vx^2 +2U|\psi|^2)\psi = (\lambda +
F\{\psi,\psi^*\})\psi,
\]
where $F\{\psi,\psi^*\}$ is the value of the functional on the
solution. On the other hand, by multiplying the above equation and
integrating we get
\[
(\lambda + F\{\psi,\psi^*\})||\psi||^2 =
\int\mathrm{d}\vx^2\psi^*(-\nabla^2 + \vx^2 +2U|\psi|^2)\psi.
\]
hence $2U = U$ and we have arrived at a contradiction unless
$U=0$. Q.E.D.

\subsection{Generalization to several order parameters }
\label{general}

The Rayleigh functional was introduced above  for a nonlinear
system described by a single (complex-valued) order parameter
$\psi(\vx)$. It can be easily generalized to the case several
order parameters.  The definition of the Rayleigh functional will
depend on the number of the conserved $l_2$-norms. This number is
determined by the type of phase invariance  of the energy
functional. We will concentrate mainly on the two broad cases: the
incoherent  and coherent coupling of the order parameters. For
simplicity, consider the case of two order parameters: $\vec{\psi}
= (\psi_1,\psi_2)$. In the case of incoherent coupling, there are
two independent constraints corresponding to the two conserved
$l_2$-norms (the normalization is defined independently for each
order parameter: $||f_l|| = 1$, $l=1,2$). On the other hand, if
the coupling is coherent, then there is only one constraint
corresponding to conservation of the total $l_2$-norm ($||f_1||^2
+||f_2||^2 = 1$). We also give an example of application of the
Rayleigh functional to a nonlinear system which does not belong to
either of the above cases (see section \ref{Num}).

An example of the incoherent coupling is the Gross-Pitaevskii
functional for a two-species mixture of degenerate  quantum gases
in an external trap below the condensation temperature (see, for
instance, Ref. \cite{BEC}):
\begin{equation}
E_{\text{nc}}\{\psi_1,\psi_1^*,\psi_2,\psi_2^*\} =
\int\mathrm{d}^3\vx\left\{ \sum_{l=1,2}N_l\left(|\nabla\psi_l|^2 +
\lambda_l^2\vx^2|\psi_l|^2\right) +
\sum_{l,m=1}^2\frac{g_{lm}}{2}N_lN_m|\psi_l|^2|\psi_m|^2 \right\}.
\label{EQ14}\end{equation}
On the other hand, the coupled-mode approximation for a
Bose-Einstein condensate in the three dimensional parabolic trap
modified in one dimension by a laser beam with creation of a
central barrier illustrates the coherent coupling:
\begin{equation}
E_{\textrm{c}}\{\psi_1,\psi_1^*,\psi_2,\psi_2^*\} =
N\int\mathrm{d}^2\vx \left\{ \sum_{l=1,2}\left(|\nabla \psi_l|^2
+\vx^2|\psi_l |^2  + \frac{g_lN}{2}|\psi_l|^4\right) -
\kappa(\psi_1\psi_2^* + \psi_2\psi_1^*) \right\}
\label{EQ15}\end{equation}
(for derivation of the coupled-mode system and the applicability
conditions consult Ref. \cite{2Ddw}). The coupling coefficient
$\kappa$ is proportional to the tunnelling rate through the
central barrier of the resulting double-well trap. It is easy to
see that functional (\ref{EQ14}) admits two independent phase
invariance transformations, $\psi_l\to e^{i\theta_l}\psi_l$, while
functional (\ref{EQ15}) admits only the simultaneous phase
invariance transformation of both variables with the same
$\theta$.  The Rayleigh functionals for the two cases
corresponding to equations (\ref{EQ14}) and (\ref{EQ15}) are
defined  as follows:
\begin{equation}
\mathcal{R}_{\textrm{nc}} =
E_{\textrm{nc}}\left\{\frac{\psi_1}{||\psi_1||},
\frac{\psi_1^*}{||\psi_1||},
\frac{\psi_2}{||\psi_2||},\frac{\psi_2^*}{||\psi_2||}\right\},
\quad
\mathcal{R}_{\textrm{c}}=
E_{\textrm{c}}\left\{\frac{\psi_1}{||\vec{\psi}||},
\frac{\psi_1^*}{||\vec{\psi}||},
\frac{\psi_2}{||\vec{\psi}||},\frac{\psi_2^*}{||\vec{\psi}||}\right\},
\label{EQ16}\end{equation}
where $||\vec{\psi}||^2 =
\int\mathrm{d}^n\vx(|\psi_1|^2+|\psi_2|^2)$.

It should be stressed  that to define the Rayleigh functional one
should employ the normalization which follows from the phase
invariance in its general form. For instance, the normalization by
the total $l_2$-norm in the incoherent coupling case would allow
one to find only the stationary solutions with the same chemical
potential for each component of the order parameter, i.e. a very
limited class of solutions.

The variational derivative  of the Rayleigh functional
$\mathcal{R}_\textrm{nc}$ (\ref{EQ16}) is derived by a mere
repetition of the steps which led to equation (\ref{EQ11}). We
have
\begin{equation}
\frac{\delta\mathcal{R}_{\textrm{nc}}}{\delta\psi_l^*} =
\frac{1}{||\psi_l||}\left\{\frac{\delta E_{\textrm{nc}}}{\delta
f_l^*}-\mathrm{Re}\left(\int\mathrm{d}^3\vx\frac{\delta
E_{\textrm{nc}}}{\delta
f_l^*}f_l^*\right)f_l\right\}\biggr|_{\vec{f} =
\left(\frac{\psi_1}{||\psi_1||},\frac{\psi_2}{||\psi_2||}\right)},
\label{EQ17}\end{equation}
where,  as usual, the $f$-variables are substituted in the energy
functional instead of $\psi$-ones, i.e.
$E_\mathrm{nc}=E_\mathrm{nc}\{f_1,f_1^*,f_2,f_2^*\}$. Now it is
easy to see that, thanks to the  equation $\mu_lf_l = \delta
E_\mathrm{nc}/\delta f^*_l$, the integral on the r.h.s. of formula
(\ref{EQ17}) coincides with the corresponding chemical potential
$\mu_l$. Thereby the extremals of the Rayleigh functional
$\mathcal{R}_\textrm{nc}$ are  stationary points of the
corresponding nonlinear system of equations.

The variational derivative of the functional
$\mathcal{R}_\textrm{c}$ (\ref{EQ16}) can be most easily obtained
from formula (\ref{EQ11}) by noticing that this Rayleigh
functional  depends on the function $\psi_l^*(\vx)$ through $f_l^*
= \psi_l^*/||\vec{\psi}||$ and also through the total norm
$||\vec{\psi}||$ in all other $f$-variables. Therefore,  each
variable of the energy functional leads to an integral term
similar to the second term in equation (\ref{EQ11}) but multiplied
by $f_l$. Hence the final result:
\begin{equation}
\frac{\delta\mathcal{R}_{\textrm{c}}}{\delta\psi_l^*} =
\frac{1}{||\vec{\psi}||}\left\{ \frac{\delta
E_{\textrm{c}}}{\delta f_l^*} -
\mathrm{Re}\left(\int\mathrm{d}^2\vx\sum_{m=1,2}\frac{\delta
E_{\textrm{c}}}{\delta
f_m^*}f^*_m\right)f_l\right\}\Biggr|_{\vec{f}=
\frac{\vec{\psi}}{||\vec{\psi}||}},
 \label{EQ18}\end{equation}
where $E_\mathrm{c}=E_\mathrm{c}\{f_1,f_1^*,f_2,f_2^*\}$. To show
that the extremals of the Rayleigh functional
$\mathcal{R}_\mathrm{c}$ are also stationary points one has to use
both  stationary equations, i.e. $\mu f_1 = \delta
E_\mathrm{nc}/\delta f^*_1$ and $\mu f_2 = \delta
E_\mathrm{nc}/\delta f^*_2$, multiplied by the complex conjugate
functions $f^*_j$. We get
\[
\int\mathrm{d}^2\vx\sum_{m=1,2}\frac{\delta E_{\textrm{c}}}{\delta
f_m^*}f^*_m = \mu\int\mathrm{d}^2\vx(|f_1|^2+|f_2|^2) = \mu.
\]

Let us now find the second variation of the Rayleigh functional at
a stationary point in the above  two cases. The derivation is
similar to that for a single order parameter: one has to use the
stationary equations and the normalization constraints for the
$f$-variables. We get:
\begin{equation}
\delta^2\mathcal{R}_{\textrm{nc}} =
\left(\delta^2E_{\textrm{nc}}\bigr|_{\delta^2 \vec{f} = 0} -
2\sum_{m=1,2}\mu_m\int\mathrm{d}^3\vx|\delta
f_m|^2\right)\Biggr|_{\vec{f} =
\left(\frac{\psi_1}{||\psi_1||},\frac{\psi_2}{||\psi_2||}\right)},
\label{EQ19}\end{equation}
\begin{equation}
\delta^2\mathcal{R}_{\textrm{c}} =
\left(\delta^2E_{\textrm{c}}\bigr|_{\delta^2 \vec{f} = 0} -
2\mu\int\mathrm{d}^2\vx\sum_{m=1,2}|\delta
f_m|^2\right)\Biggr|_{\vec{f} =
\frac{\vec{\psi}}{||\vec{\psi}||}}.
\label{EQ20}\end{equation}
In both cases the second variation of the Rayleigh functional
coincides with that of the Lagrange modified energy functional
considered in the space of normalized functions. Therefore, the
minima of the Lagrange modified energy functional (for the fixed
values of the conserved $l_2$-norms) are also minima of the
Rayleigh functional and vice versa.

Concluding this section we note that by introducing  the concept
of  Rayleigh functional we have converted the task of finding the
stationary points to an unconstrained optimization problem which
also distinguishes between the local minima and the saddle points.
In the next section we give examples of application of this
optimization method  to the numerical search for the stationary
states of Bose-Einstein condesates.

\section{Numerical search for  stationary points}
\label{Num}

By analogy with the use of the Rayleigh quotient in  the numerical
search for eigenfunctions, the nonlinear optimization can be based
on the Rayleigh functional. Leaving a mathematically rigorous
analysis for future research, let us nevertheless  give an
heuristic argument. The Rayleigh functional is a compound
functional: the energy functional depends on $f(\vx)$ and
$f^*(\vx)$, while these are functionals of $\psi(\vx)$ and
$\psi^*(\vx)$. The numerical nonlinear optimization of a
functional reduces to that of a smooth function (though of a large
number of variables).  But a continuous function on a
finite-dimensional sphere (which in the numerics plays the role of
the variable $f(\vx)$) always has a minimum. Since the second
variation of the Rayleigh functional is actually the second
variation of the Lagrange modified energy functional, evaluated in
the space of normalized functions, the optimization method based
on the Rayleigh functional has the capacity to return the absolute
minimum. Generally,  it converges to a local minimum
\footnote{However, some particular choices of the initial profile
for the order parameter may result in convergence to the nearest
saddle point as if it were a minimum.}.

An additional argument in favor of the nonlinear optimization
based on the Rayleigh functional  is provided already by the
imaginary time relaxation method -- a widely used approach in the
computational physics. Indeed, it is precisely the Rayleigh
functional which is minimized in this method. Let us recall the
usual formulation given by equation (\ref{EQ4}) of section
\ref{intro}. First of all, this formulation does not reflect the
total change of $\psi$, since after each numerical step  one
performs the direct normalization of the updated function. This
means that the variational derivative $\delta E/\delta f^*$ does
not tend to zero, since the norm of the order parameter is not
forced to approach a constant value as $\tau\to\infty$. However,
one can reformulate the relaxation method for the variable
$f(\vx,\tau)= \psi(\vx,\tau)/||\psi(\vx,\tau)||$ which has a fixed
$l_2$-norm. We have
\[
\frac{\partial f}{\partial \tau} = \frac{\partial}{\partial
\tau}\left(\frac{\psi}{||\psi||}\right) =
\frac{1}{||\psi||}\left\{\frac{\partial\psi}{\partial
\tau}-\frac{\psi}{2||\psi||^2}\int\mathrm{d}^n\vx
\left(\psi\frac{\partial\psi^*}{\partial
\tau}+\psi^*\frac{\partial\psi}{\partial \tau}\right)\right\}
\]
\begin{equation}
= - \frac{\delta \mathcal{R}\{\psi,\psi^*\}}{\delta
\psi^*}\biggr|_{f = \frac{\psi}{||\psi||}},
\label{EQ21}\end{equation}
where we have used equation (\ref{EQ4}) to substitute for
$\partial\psi/\partial\tau$. We see that if the imaginary time
relaxation  converges, i.e. if  ${\partial f}/{\partial
\tau}\to0$, then   it converges necessarily to an extremal of the
Rayleigh functional.

Equation (\ref{EQ21}), besides revealing  a new side of the imaginary time
evolution method,   manifests its strong drawback: the order parameter evolves
continuously and, hence, quite slowly. But, since it is actually a way to minimize
the Rayleigh functional, a discrete minimization scheme can be adopted with the
order parameter performing a series of finite steps to a local minimum.  We have
naturally arrived at the steepest descent formulation:
\begin{equation}
\psi^{(k+1)} = \psi^{(k)} + D^{(k)},\quad D^{(k)} \equiv
-\frac{1}{\alpha_k}\frac{\delta\mathcal{R}\{\psi^{(k)},\psi^*{}^{(k)}\}}{\delta
\psi^*{}^{(k)}}
\label{EQ23}\end{equation}
where the parameter $\alpha_k$ is selected by some line-search
algorithm. More advanced gradient methods, for example, the
conjugate gradient method, can be employed for the finite-step
minimization. However, we have found that the  steepest descent
algorithm performs very well in the nonlinear optimization based
on the Rayleigh functional.

In selecting the step size $1/\alpha_k$  we have adopted the
Barzilai-Borwein two-point method  \cite{BB} (see also
Ref.~\cite{onBB1,onBB2}). There are two closely related
Barzilai-Borwein methods:
\begin{subequations}
\label{EQ24}
\begin{eqnarray}
\alpha^{(1)}_k &=& \frac{\mathrm{Re}\int\mathrm{d}^n\vx
D^*{}^{(k-1)}\left(\frac{\delta\mathcal{R}^{(k)}}{\delta
{\psi^*}{}^{(k)}} -\frac{\delta\mathcal{R}^{(k-1)}}{\delta
\psi^*{}^{(k-1)}}\right)}{\int\mathrm{d}^n\vx |D^{(k-1)}|^2},
\label{EQ24a}\\
\alpha^{(2)}_k &=& \frac{\int\mathrm{d}^n\vx
\left|\frac{\delta\mathcal{R}^{(k)}}{\delta {\psi^*}{}^{(k)}}
-\frac{\delta\mathcal{R}^{(k-1)}}{\delta
\psi^*{}^{(k-1)}}\right|^2} {\mathrm{Re}\int\mathrm{d}^n\vx
D^*{}^{(k-1)}\left(\frac{\delta\mathcal{R}^{(k)}}{\delta
{\psi^*}{}^{(k)}} -\frac{\delta\mathcal{R}^{(k-1)}}{\delta
\psi^*{}^{(k-1)}}  \right)}.
\label{EQ24b}\end{eqnarray}
\end{subequations}
Both methods are fairly equivalent in terms of performance. The
generalization of the numerical scheme (\ref{EQ23})-(\ref{EQ24})
to several order parameters is given by the substitution of the
scalar function $\psi(\vx)$ by a vectorial one $\vec{\psi}(\vx)$
defined similar to the vectorial function $\vec{f}(\vx)$ used in
formulae (\ref{EQ17}) and (\ref{EQ18}) (in the vector case, the
inner product in equations (\ref{EQ24a}) and (\ref{EQ24b})
contains, besides the integral, the Hermitian inner product in the
finite-dimensional target vector space).

In the numerical implementation we have used the spectral
collocation  method for the spatial grid  based on the Fourier
modes (see Refs. \cite{PS1,PS2,PS3}). The accuracy of the obtained
solution was checked by  direct numerical substitution into the
governing equations. In all cases a rapid convergence to the
stationary point was observed. We have cut off the allowed values
of $1/\alpha$  by a small positive number (in our case
$1/\alpha>10^{-8}$) to have the iteration step always in the
direction of decrease of the Rayleigh functional.

\subsection*{Examples of numerical search for the stationary states }

\textit{A. Incoherent coupling}. Our  first example of the
numerical optimization based on the Rayleigh functional is the
calculation of the ground state in the two-species Bose-Einstein
condensate of Rubidium in the quasi two-dimensional geometry (the
pancake trap). The derivation of the governing equations and
further discussion can be found in Ref. \cite{MixBEC}. The energy
functional for the two-species Bose-Einstein condensate is given
by the two-dimensional version of  equation (\ref{EQ14}). The
parameters $\lambda_1$ and $\lambda_2$ account for the different
Lande magnetic factors. For the two isotopes of Rubidium we have:
$\lambda^2_1 = 8/7$ (for ${}^{85}$Rb) and $\lambda^2_2 = 6/7$ (for
${}^{87}$Rb), while the interaction coefficients are $g_{11} =
-0.0219$, $g_{22} = 0.0068$, and $g_{12} = 0.012$. The variational
derivative is given by equation (\ref{EQ17}). To simplify the
task, we note that the first term of the variational derivatives
is the r.h.s of the stationary Euler-Lagrange equations written
for the normalized functions:
\begin{subequations}
\label{EQincoh}
\begin{eqnarray}
\mu_1 f_1 &=&  N_1\left(-\nabla^2 f_1 + \lambda_1^2 \vec{x}{\,}^2
f_1\right) +
\left(g_{11}N_1^2|f_1|^2  +  g_{12}N_1N_2|f_2|^2\right)f_1, \\
\mu_2 f_2 &=&  N_2\left(-\nabla^2 f_2 + \lambda_2^2 \vec{x}{\,}^2
f_2 \right) + \left(g_{12}N_1N_2|f_1|^2 +
g_{22}N_2^2|f_2|^2\right)f_2.
\end{eqnarray}
\end{subequations}
Here the chemical potentials $\mu_j$ are defined as follows:
$\mu_j = \mu_{N_j} N_j$, $j=1,2$, with $\mu_{N_j}$ being the
chemical potential for the order parameter $\psi_j$ normalized to
the number of atoms $N_j$.

\begin{figure}[hp]
\includegraphics{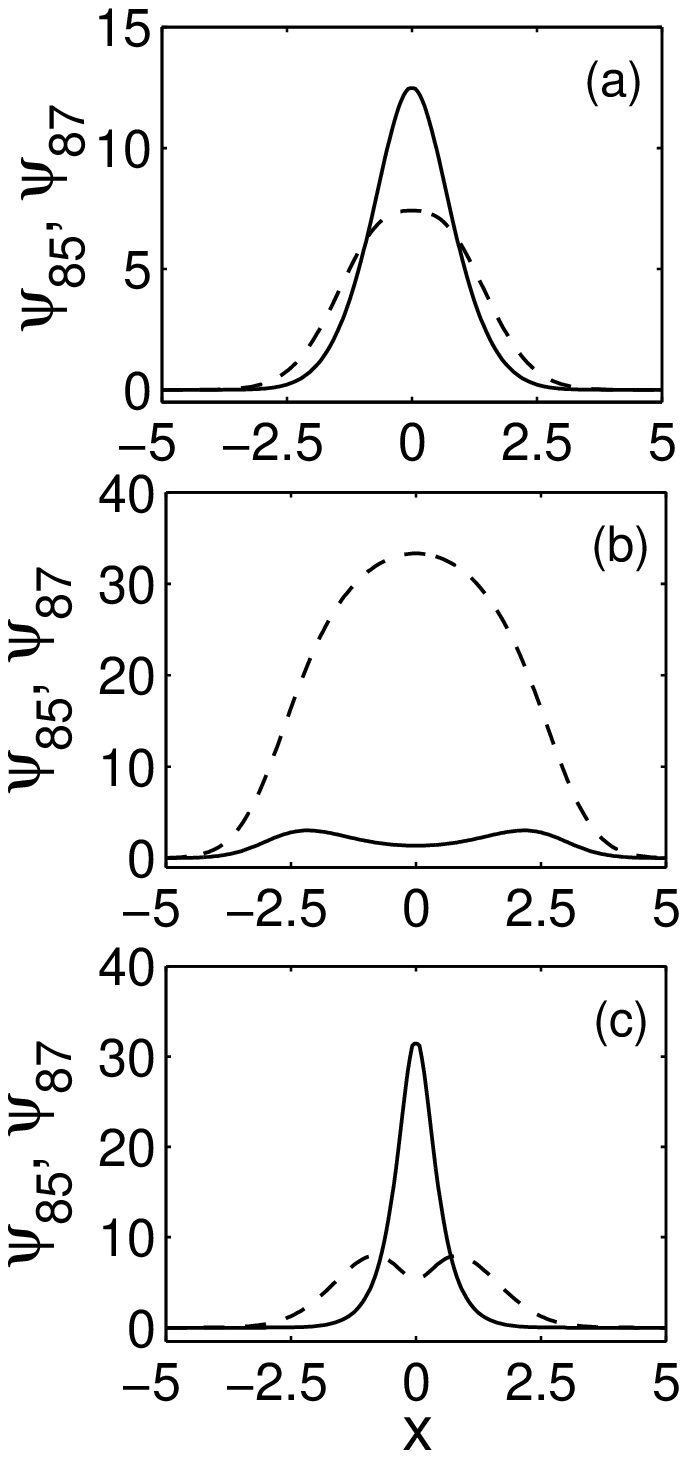}
\caption{\label{FG1} The axially symmetric ground state of the
two-isotope Bose-Einstein condensate mixture  of Rubidium. The
dashed and dotted lines give the order parameters (normalized to
the actual number of atoms) of ${}^{85}$Rb and ${}^{87}$Rb
isotopes, respectively. Panel (a) $N_{85}=N_{87}=300$, $\mu_{85} =
1.18$ and $\mu_{87} = 2.67$; panel (b) $N_{85}=200$,
$N_{87}=15000$, $\mu_{85} = 11.69$ and $\mu_{87} = 7.827$; panel
(c) $N_{85}=N_{87}=500$, $\mu_{85} = -3.48$ and $\mu_{87} = 3.08$
(in dimensionless units). }
\end{figure}

\begin{figure}[hp]
\includegraphics{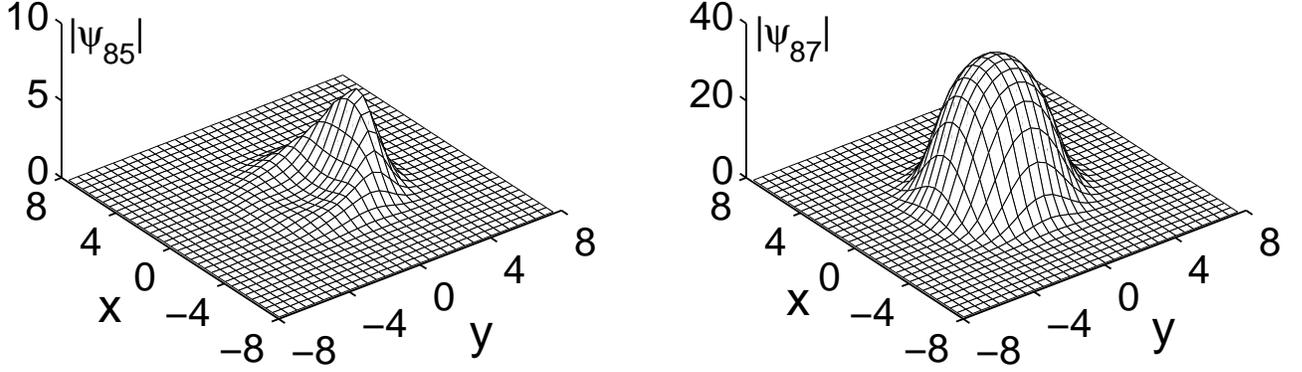}
\caption{\label{FG2} The symmetry breaking ground state of the
 Bose-Einstein condensate mixture of the isotopes of Rubidium. The left
and right panels give the shapes of the order parameters of
${}^{85}$Rb and ${}^{87}$Rb isotopes, respectively. Here $N_{85} =
150$, $N_{87} = 20000$, $\mu_{85} = 13.19$ and $\mu_{87} = 8.91$
(in dimensionless units). }
\end{figure}

A nontrivial feature of the two-species condensate is that the
ground state suffers from the symmetry breaking transformation, if
for instance, for a fixed number of atoms of the ${}^{85}$Rb
isotope the number of atoms of the ${}^{87}$Rb isotope increases
\cite{MixBEC}. In figure \ref{FG1} we show the symmetry preserved
(axially symmetric) ground states, while the symmetry-broken one
is illustrated in figure \ref{FG2}. The axially symmetric states
were found by the optimization method projected on the space of
axially symmetric functions and formulated in polar coordinates.
The polar radius grid contained 128 points, while the
two-dimensional grid was $64\times64$ (we have used the
$32\times32$-grid to produce figure \ref{FG2}). The initial
condition in both cases was the Gaussian profile (modulated by a
symmetry-breaking perturbation in case of figure \ref{FG2}). Our
numerical simulations were performed in the MATLAB. The typical
number of the iterations to reach the $10^{-10}$ convergence was
between 400-1000 (on a personal computer with an AMD
1Ghz-processor  the iterations have taken up to 5 seconds in the
axially-symmetric case and up to 30 seconds in the 2D case).

\textit{B. Coherent coupling}. Consider the following functional
\[
E_{\textrm{c}}\{\psi_1,\psi_1^*,\psi_2,\psi_2^*\} =
\int\mathrm{d}^2\vx  \bigl\{|\nabla \psi_1|^2 + |\nabla \psi_2|^2
+\vx^2\left(|\psi_1 |^2 +|\psi_2|^2\right) + \varepsilon|\psi_2|^2
\]
\begin{equation}
- \frac{N}{2}|\psi_1|^4 +  \frac{aN}{2}|\psi_2|^4 -
\kappa(\psi_1\psi_2^* + \psi_2\psi_1^*)\bigr\},
\label{EQ25}\end{equation}
which appears in the description of the stationary states of a
pair of repulsive and attractive ($a>0$) two-dimensional
condensates trapped in an asymmetric double-well potential
\cite{2Ddw} (in this case $N$ is proportional to the actual number
of condensate atoms with the coefficient of proportionality of the
order $10^2-10^3$).  Here $\varepsilon$ is the zero-point energy
difference between the two wells of the external trap and $N$ is
the total number of atoms. The Euler-Lagrange equations for the
energy functional (\ref{EQ25}) read:
\begin{subequations}
\label{EQcoh}
\begin{eqnarray}
\mu f_1 &=& -\nabla^2 f_1 + \vec{x}{\,}^2 f_1
-\kappa f_2 - N|f_1|^2f_1, \\
\mu f_2 &=&  -\nabla^2 f_2 +  \varepsilon f_2 + \vec{x}{\,}^2 f_2
-\kappa f_1  +  aN|f_2|^2f_2,
\end{eqnarray}
\end{subequations}
where $\mu$ coincides with the chemical potential for the order
parameter normalized to the number of atoms $N$.

\begin{figure}[hp]
\includegraphics{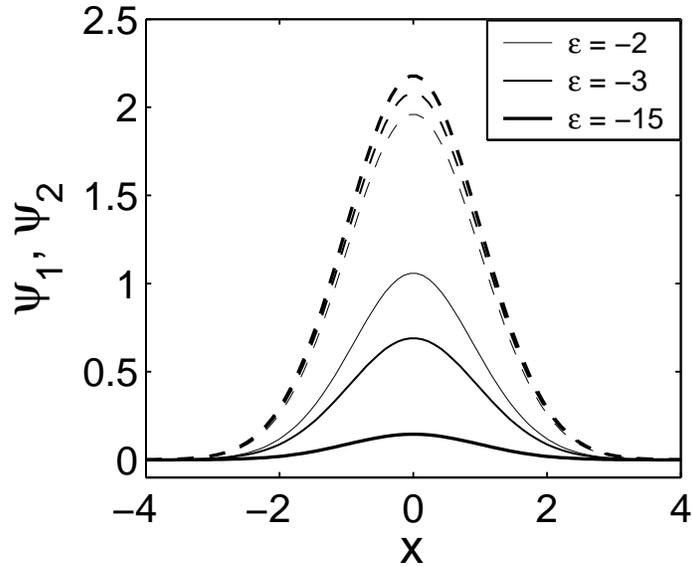}
\caption{\label{FG3} Deformation of the ground state of two
Bose-Einstein condensates  trapped in an asymmetric double-well
potential with variation of the zero-point energy difference. The
picture corresponds to the energy functional (\ref{EQ25}).  Here
$a=0$, $\kappa=1$ and $N = 15$, the solid line gives $\psi_1$,
while the dashed one $\psi_2$. }
\end{figure}

As in the previous example of the incoherent coupling, we have
observed a rapid convergence of the  iterations  to the ground
state solution. The numerical minimization was performed in polar
coordinates. The characteristic times to reach the deviation of
about $10^{-10}$ are the same as in the previous example
(formulated in polar coordinates). To verify that the solutions
obtained with the use of the Rayleigh functional method are indeed
the ground states, we have used another numerical approach which
enables one to find all possible stationary states (consult for
more details Refs. \cite{2Ddw,PhysD}). The deformation of the
ground state with the variation of the zero-point energy
difference is illustrated in figure \ref{FG3}, where the
condensate of vanishing atomic interaction, $a=0$, is
tunnel-coupled to an attractive condensate.

\textit{C. Arbitrary coupling}. The minimization method based on
the Rayleigh functional can be applied to systems of equations
coupled by a different way than the coherent and incoherent
coupling. As an example, we consider here the Bose-Einstein
condensate in a pancake trap (in the $\vec{r}_\perp = (x,y)$
plane), but in contrast to the above considered cases, we take
into account the transverse dimension ($z$) by using the Gaussian
variational ansatz with the $\vec{r}_\perp$-dependent parameters,
\begin{equation}
\psi  =
\frac{1}{\pi^{1/4}w^{1/2}(\vec{r}_\perp,t)}\exp\left(-\frac{z^2}
{2w^2(\vec{r}_\perp,t)}\right)
f(\vec{r}_\perp,t)e^{i\alpha(\vec{r}_\perp,t)z^2/2},
\label{EQN1}\end{equation}
for the order parameter in the Gross-Pitaevskii functional
(\ref{EQ2}). A similar ansatz was already used with success for
the condensate in a cigar shaped trap \cite{1D}. Here the
parameter $\alpha z^2/2$ accounts for the transverse motion of the
condensate.  We are interested in the stationary solutions which
correspond to $\alpha=0$ and are given as $f = e^{-i\mu
t/\hbar}U(\vec{r}_\perp)$ and $w = w(\vec{r}_\perp)$. Integrating
over the transverse coordinate in the Gross-Pitaevskii functional
(\ref{EQ2})  we get the energy functional for the stationary
states:
\begin{equation}
E_{2D} = \int\mathrm{d}^2\vec{r}_\perp \left\{
\frac{\hbar^2}{2m}|\nabla_\perp U|^2 + V(\vec{r}_\perp)|U|^2 +
\frac{gN}{2\sqrt{2\pi}}\frac{|U|^4}{w} + \frac{\hbar^2}{4m}\left(
\frac{(\nabla_\perp w)^2}{w^2} + \frac{1}{w^2} +
\frac{w^2}{a_z^4}\right)|U|^2\right\}
\label{EQN2}\end{equation}
Here  $a_z = (\hbar/m\omega_z)^{1/2}$ the transverse oscillator
length, $V(\vec{r}_\perp) = m\omega^2_\perp{\vec{r}_\perp}^{\,
2}/2$ is the external potential in 2D. Note that $w \to a_z$ as
$|\vec{r}_\perp|\to \infty$. The pancake geometry corresponds to
$\lambda = \omega_z/\omega_\perp\gg1$.  The equations satisfied by
the stationary states read:
\begin{equation}
\mu U +\frac{\hbar^2}{2m}\nabla^2_\perp U - V(\vec{r}_\perp)U -
\frac{g}{\sqrt{2\pi}\,w}|U|^2U - \frac{\hbar^2}{4m}\left(
\frac{(\nabla_\perp w)^2}{w^2} + \frac{1}{w^2} +
\frac{w^2}{a_z^4}\right)U =0,
\label{EQN3}\end{equation}
\begin{equation}
w\nabla_\perp\left( \frac{\nabla_\perp w}{w^2}|U|^2\right)
+\left(\frac{(\nabla_\perp w)^2}{w^2} + \frac{1}{w^2}
-\frac{w^2}{a_z^4}\right)|U|^2 + \frac{2\sqrt{2\pi}a_s}{w}|U|^4=0,
\label{EQN4}\end{equation}
where $a_s$ is the $s$-wave scattering length.

Assuming that the transverse width $w$ is of the order of $a_z$
(which is supported by the numerics) and estimating the operator
$\nabla_\perp \sim 1/a_\perp$ we see that the term $U(\nabla_\perp
w)^2/(w^2)$ is much less than $(\frac{1}{w^2} +
\frac{w^2}{a_z^4})U$ for $\lambda\gg1$. If this small term is
omitted from the energy functional (\ref{EQN2}), the equations for
the stationary state become equivalent to the two-dimensional
non-polynomial NLS equation (2D NPSE) of Ref.~\cite{Sal}. (To
compare with the latter reference, though, one should take into
account that the coefficient 1/2 is missing there at
$(\frac{1}{w^2} + \frac{w^2}{a_z^4})U$ in the equation which is
equivalent to equation (\ref{EQN3})).

The essential difference of the energy functional given by
equation (\ref{EQN2}) from the above considered energy functionals
lies in the fact that it has two variables of variation, $U$ and
$w$, but only one phase invariance exists and is associated with
the normalization of $U$. However, the minimization method based
on the Rayleigh functional defined with the substitution $f \to
U/||U||$ into the energy functional (\ref{EQN2}) works in this
case also. To explain this, we note that one can think of the
``chemical potential'' corresponding to the variable $w$ as being
equal to zero, thus the variation of the Rayleigh functional with
respect to $w$ coincides with that of the energy functional.

The solutions of equations (\ref{EQN3}) and (\ref{EQN4}) were
searched for in polar coordinates. As the system similar to
(\ref{EQN3})-(\ref{EQN4}) was proposed in Ref. \cite{Sal} as an
improvement to the NLS equation in two dimensions,  we aimed to
find out when does one need to use this complicated system for the
pancake condensate instead of the much simpler 2D NLS equation? We
have found that only in the study of collapse instability for
$g<0$ the system (\ref{EQN3})-(\ref{EQN4}) gives visible
improvement over the 2D NLS equation. In all other cases, if one
is not interested in the transverse profile of the condensate, the
2D NLS equation gives satisfactory results for the stationary
states. (This, however, may not be so for the arbitrary
time-dependent evolution, which is not captured by the 2D NPSE
system of Ref. \cite{Sal} since their $\alpha =0$ identically.)

\begin{figure}[hp]
\includegraphics{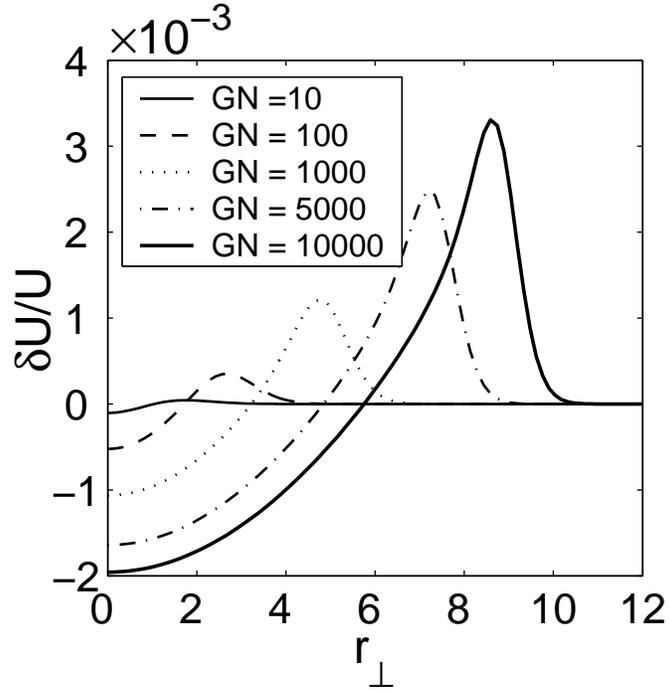}
\caption{\label{FG4} The relative difference between the solutions
of the 2D NLS equation and the system (\ref{EQN3})-(\ref{EQN4}).
Here $\lambda = 100$. }
\end{figure}

\begin{figure}[hp]
\includegraphics{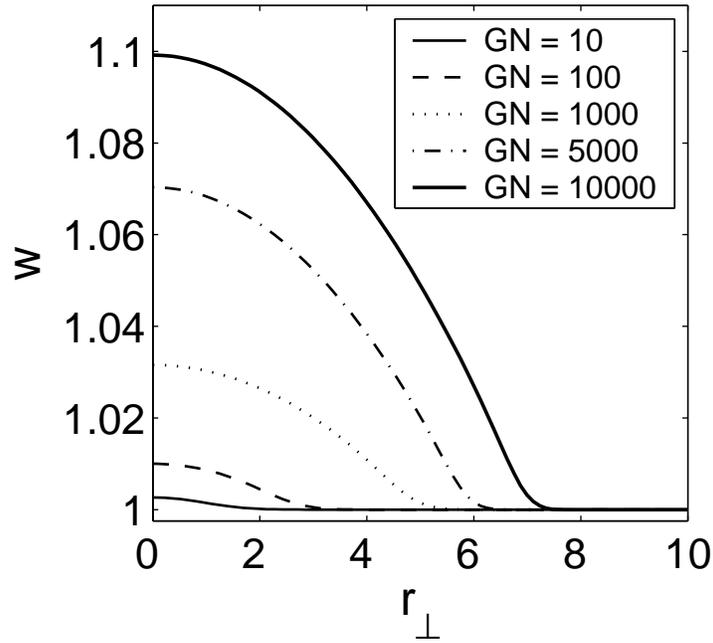}
\caption{\label{FG5} The transverse width of the condensate given
by the system (\ref{EQN3})-(\ref{EQN4}). Here $\lambda = 100$. }
\end{figure}

To illustrate this, in fig.~\ref{FG4} we give the relative
difference between the solutions $U(r_\perp)$ of the system
(\ref{EQN3})-(\ref{EQN4}) and the corresponding 2D NLS equation
for several values of the nonlinearity strength in the case of
repulsive condensate ($g>0$). The solution is given in the units
of $1/a_\perp$, $a_\perp= (\hbar/m\omega_\perp)^{1/2}$, the radius
$r_\perp$ in terms of $a_\perp$, and $\mu$ in terms of
$\hbar\omega_\perp/2$. The dimensionless nonlinearity strength is
defined as $G = 4\sqrt{2\pi}a_s/a_z$. We have set $\lambda=100$
which corresponds to the ratio $a_\perp/a_z = 10$. The transverse
width $w$, given in terms of $a_z$, is shown in fig. \ref{FG5}.
The maximal difference from $a_z$ is of the order of 10\%.

\begin{figure}[hp]
\includegraphics{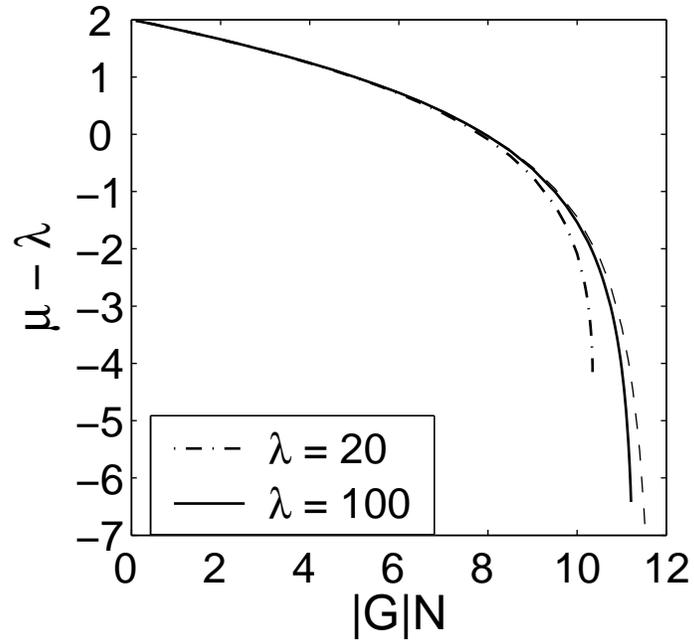}
\caption{\label{FG6} The dependence of the chemical potential on
nonlinearity for the  ground state of the system
(\ref{EQN3})-(\ref{EQN4}). The dashed line gives the same for the
2D NLS equation with the parabolic potential. }
\end{figure}

Though in the case of attractive condensate the relative
difference between the solutions of the system
(\ref{EQN3})-(\ref{EQN4}) and the 2D NLS equation is also small,
the critical value for collapse differs noticeably. The critical
value of the number of atoms is defined by $\partial \mu/\partial
N = \infty$, i. e. it is the endpoint coordinate of the curves in
fig. \ref{FG6}. The system (\ref{EQN3})-(\ref{EQN4}) is an
improvement over the NLS equation in this case, since the critical
value differs by only 2\% for $\lambda \ge 20$ from the actual
value of the 3D condensate obtained in Ref. \cite{Collapse3D}. The
2D NLS value is achieved as the asymptotic limit when $\lambda\to
\infty$.

We have argued that the minimization based on the Rayleigh
functional is always suitable for obtaining the ground state
solutions. However, the possibility of obtaining excited states
was not excluded at all. In contrast to the linear eigenvalue
problem of section \ref{Ray} in the nonlinear case there is no
general approach which would guarantee computation of all excited
states, though there are some advances in this direction
\cite{SobolGrad}. However, a proper selection of the initial
condition for minimization may lead to convergence to the
``nearest'' excited state. We illustrate this on the multi-vortex
solutions of a nonlocal Gross-Pitaevskii equation, which in
dimensionless form, in the frame of reference rotating with the
frequency $\Omega$, reads (see the details in Ref. \cite{VORT})
\begin{equation}
i\partial_t \Psi = -\nabla^2\Psi + \vx^2\Psi - \Omega L_z\Psi +
g\Psi\int\mathrm{d}{\vx^\prime}^2
K(|\vx-\vx^\prime|)|\Psi(\vx^\prime)|^2.
\label{GP3D}\end{equation}
Here $K  = \frac{1}{2\pi a^2}K_0\left(\frac{r}{a}\right)$,
$K_0(z)$ is the Macdonald function, $a$ is the effective range of
atomic interaction, $L_z$ is the angular momentum projection on
the $z$-axis: $L_z = -i(x\partial_y - y\partial_x)=
-i\partial_\theta$, where $\theta$ is the polar angle, and
$\Omega$ is the rotation frequency. For simplicity, we consider
the periodic boundary conditions along the $z$-axis, $\Psi(z+d) =
\Psi(z)$.

We searched for new non-axial vortex solutions, involving
combinations of vortices and antivortices, which represent the
excited states of the system, since in the nonlocal GP equation
(\ref{GP3D}), similar as the  in local one, the stationary vortex
solutions involving combinations of vortices and antivortices are
never the energy minimizers. The energy is minimized by Tkachenko
lattices comprised of vortices only \cite{VORT}.

\begin{figure}
\begin{center}
\includegraphics{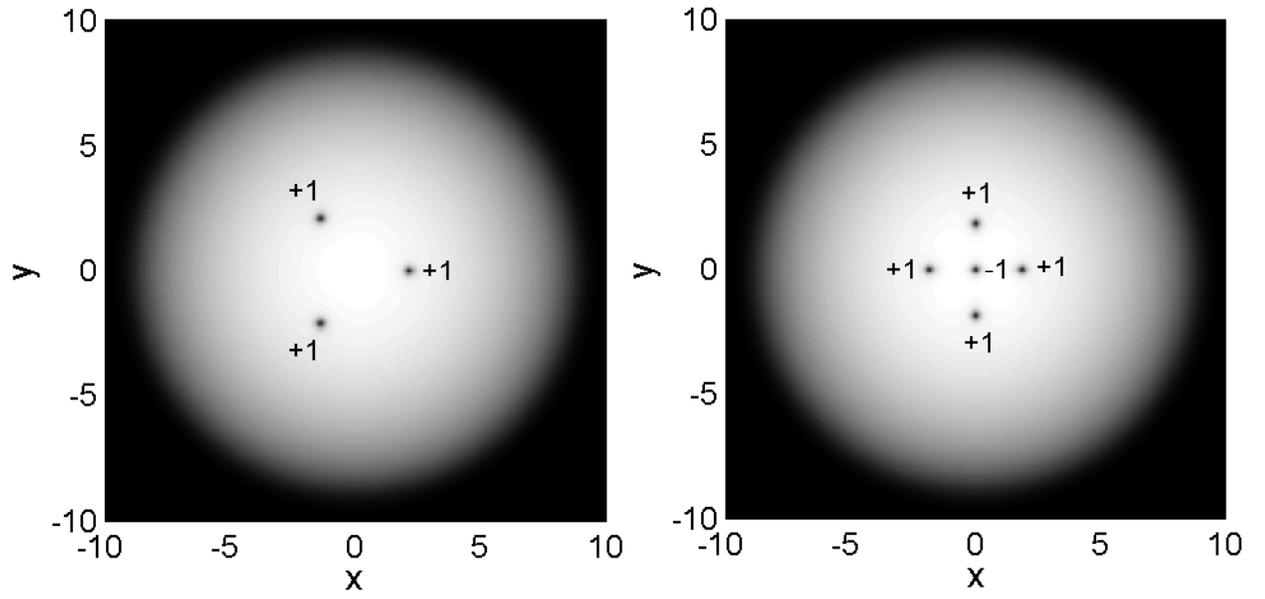}
\caption{\label{FG7}The stationary vortex solutions for $a=0.052$,
$g=10^4$ and $\Omega = 0.38$. The left panel shows the 3-vortex
solution and  the right one -- the 5-vortex solution comprised of
4 vortices and 1 antivortex (in the center). }
\end{center}
\end{figure}

\begin{figure}
\begin{center}
\includegraphics{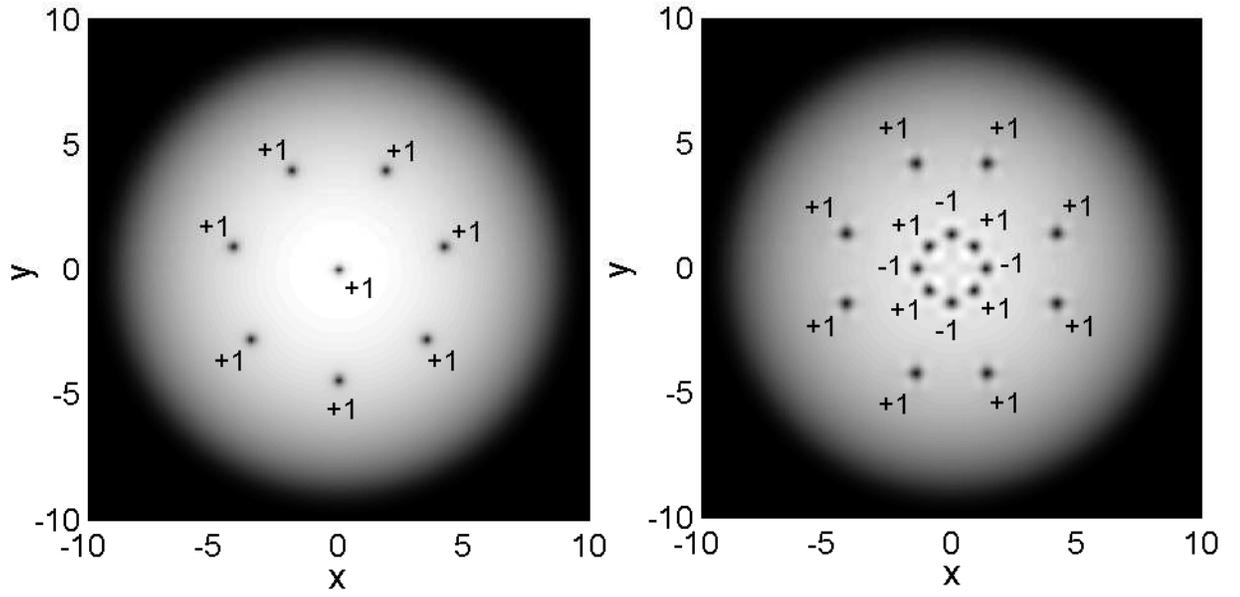}
\caption{\label{FG8}The stationary vortex solutions for $a=0.052$,
$g=10^4$ and $\Omega = 0.5$. The left panel shows the 8-vortex
solution and the right one -- the 16-vortex solution comprised of
12 vortices and 4 antivortices. }
\end{center}
\end{figure}

In figures \ref{FG7} and \ref{FG8} we give two examples of excited
states comprised of vortex and anti-vortex combinations (the right
panel) and the respective energy minimizers (the left panel).
These solutions were found by using the initial conditions with
the phase resembling that of the multi-vortex solution in quest.

\section{Rayleigh functional in the analytical approach}
\label{criticalNLS}

The ground state solution of a nonlinear partial differential
equation is degenerate if fixing the $l_2$-norm does not select a
unique chemical potential.  In the critical case, when there are
infinitely many stationary points with equal energy and the same
$l_2$-norm, the nonlinear optimization method based on the
Rayleigh functional may not converge at all.

To clarify this let us consider the critical NLS equation. The
ground state of the focusing critical NLS equation is infinitely
degenerate.  It is well-known (see, for instance, Refs.
\cite{NLS,Collapse2D}) that the two-dimensional critical NLS
equation has a family of soliton solutions having the same
$l_2$-norm $||\psi||^2 \approx 11.69$ and a specific member $\psi
= e^{it}R(\vx)$ satisfying the stationary equation
\begin{equation}
\nabla^2 R + R^3 -R = 0
\label{EQ27}\end{equation}
is called the Townes soliton. The stationary NLS equation
\begin{equation}
\mu\psi + \nabla^2\psi + |\psi|^2\psi  = 0
\label{EQ28}\end{equation}
admits the scale invariance given by $\psi(\vx)\to k\psi(k\vx)$
and $\mu\to k^2\mu$. Easy to see that, in two spatial dimensions,
the scale invariance preserves the $l_2$-norm of the solution,
while the energy of the stationary solution is exactly zero (see
below). This is precisely why one cannot obtain the Townes soliton
by the nonlinear optimization method based on the Rayleigh
functional -- the method fails to converge.

This drawback of the Rayleigh functional in the numerical approach
turns out into an advantage for the analytical analysis. Indeed,
for the  critical NLS equation it immediately leads to the
well-known identity \cite{Collapse2D}
\begin{equation}
-\mu\int\mathrm{d}^2\vx|\psi|^2 =
\int\mathrm{d}^2\vx|\nabla\psi|^2 =
\frac{1}{2}\int\mathrm{d}^2\vx|\psi|^4,
\label{EQ29}\end{equation}
which is satisfied by any stationary solution. Let us show this.
First of all, it is easy to see that the energy functional of the
NLS equation $E=\int\mathrm{d}^2\vx(|\nabla\psi|^2 - |\psi|^4/2)$
has a scale invariance property: $E \to k^2E$ for $\psi(t,\vx)\to
k\psi(k^2t,k\vx)$. The invariance property is transferred to the
Rayleigh functional: $\mathcal{R} \to k^2\mathcal{R}$.
Differentiation of the latter with respect to the scale invariance
factor $k$ gives $\mathcal{R} = 0$ at the stationary point. Hence,
the second equality in equation (\ref{EQ29}) follows. The first
equality is derived by integrating the stationary equation
multiplied by $\psi^*$ and using the already proven equality.

In the above derivation we have explicitly related identity
(\ref{EQ29}) to the scale invariance property. The  scale
invariance is responsible for the exact balance of energies in any
stationary solution of the critical NLS equation: the kinetic
energy is balanced by the energy due to the self-interaction.

The identity (\ref{EQ29}) is well-known. However, the point we try
to make here that its relation to the critical scaling makes such
an identity universal, i.e. it appears in connection to any
nonlinear equation for which  the energy functional has a scaling
invariance for a family of solutions with the same $l_2$-norm.
Such is also  the one-dimensional critical NLS equation (in the
stationary form)
\begin{equation}
\mu \psi + \frac{\mathrm{d^2\psi}}{\mathrm{d}x^2} + |\psi|^4\psi =
0.
\label{EQ30}\end{equation}
In  this case the scale invariance is $\psi(x) \to
\sqrt{k}\psi(kx)$ and $\mu \to k^2\mu$ leads to the energy scaling
$E \to k^2 E$, with
$E=\int\mathrm{d}x(|\mathrm{d}\psi/\mathrm{d}x|^2 - |\psi|^6/3)$,
while the $l_2$-norm $N = \int\mathrm{d}x |\psi|^2$ remains
unchanged. Hence the Rayleigh functional scales as follows
$\mathcal{R} \to k^2\mathcal{R}$ and, by repetition of the above
arguments, we get a similar identity satisfied by the family of
stationary solutions:
\begin{equation}
-\mu\int\mathrm{d}x|\psi|^2 =
\int\mathrm{d}x\left|\frac{\mathrm{d}\psi}{\mathrm{d}x}\right |^2
= \frac{1}{3}\int\mathrm{d}x|\psi|^6.
\label{EQ31}\end{equation}

Note that one cannot use the energy functional in the above
argument: the stationary solution is \textit{not} an extremal of
the energy functional. Using the Lagrange modified functional will
not help either due to the explicit appearance of $\mu$. The use
of the Rayleigh functional is indispensable  in this short
derivation.

\section{Conclusion}
\label{Concl}

Numerical search for  stationary points of nonlinear partial
differential equations is a difficult problem. To tackle such a
problem one is left to try various methods and choose the one
which gives a better performance and accuracy.  In this paper we
have proposed a new numerical method, the nonlinear optimization
method based on the Rayleigh functional. This method is a natural
generalization of  Poincar\'e's minmax principle for  linear
equations, formulated with the use of the Rayleigh quotient.

It turns out that the imaginary time relaxation method, a widely
used method in the computational physics, is just a special case
of nonlinear optimization based on the Rayleigh functional. We
have used the gradient scheme for minimization of the Rayleigh
functional, but this is not essential: one can use the
minimization schemes which do not require use of the gradient.

The simplicity of the Rayleigh functional and its universality for
nonlinear equations is one of the advantages of the method.
Moreover, the method can distinguish between the local minima and
saddle points, since the second variation of the Rayleigh
functional is equal to the second variation of the Lagrange
modified energy functional, if the latter is evaluated in the
space of normalized functions. The Rayleigh functional takes care
of the normalization constraint of the stationary solution. Other
constraints, however, have to be treated in the usual way.

There are, however, some exceptional cases when the nonlinear
optimization based on the Rayleigh functional may fail. The
principal cause of the failure is the infinite degeneracy of the
ground state solution.  Still, a failure in the numerics is
partially compensated by the fact that the very degeneracy allows
one to get an important analytical insight relating the scale
invariance and an identity satisfied by the stationary solutions
in the critical case.

\section*{Acknowledgements}
This work was supported by  CNPq and FAPEAL of Brazil. The initial
part of this work was done during the author's visit of the
Instituto de F{i}sica Te\'{o}rica, Universidade Estadual Paulista,
in S\~ao Paulo, Brazil, which was supported by the FAPESP. We are
grateful to Jianke Yang for helpful suggestions.

\newpage
\section{Figures}

\end{document}